\documentclass[pra,twocolumn,showpacs,floats,eqsecnum,amsmath,amssymb]{revtex4}

\def\al{&\!\!\!\!}
\def\f{\frac}

\begin{document}

\title {Minimalist's Electromagnetism- \\  Different Axioms and Different Insight}
\author {Y. Sobouti}
\affiliation{Institute for Advanced Studies in Basic Sciences (IASBS)\\
Gava Zang, Zanjan 45137-66731, Iran \\
     email: sobouti@iasbs.ac.ir}

\begin{abstract}
That  the speed of light is a universal constant is a logical
consequence of Maxwell's equations.
Here we show  the converse is also true. Electromagnetism (EM) and electrodynamics (ED), in  all details,  can be derived from two simple assumptions:
i)  the speed of light is a universal constant  and,
ii) the common observations that  there are the so-called charged particles that interact with each other.
  Conventional EM and ED are observation based.
The proposed alternative spares all those  observational foundations, only to reintroduce them
as theoretically derived and empiricism-free laws of Nature.
There are merits to simplicity.  
For instance, when one learns that Poisson's equation  emerges as a corollary  of the formalism, one immediately concludes  that Coulomb's $1/r^2$ law of force is exact.   Or, if it turns out that   $\nabla.\mathbf{B}=0$ follows from the theory, then non-existence of  (at least  classical) magnetic monopoles will be an exact law of Nature.
The list is longer than the these two examples.

\end{abstract}


\maketitle
\section{Introduction}
Electromagnetism as we know it today is founded on the laboratory findings 
of Coulomb, Ampere, Faraday, and many other experimenters.
Encapsulated in Maxwell's equations, EM is a robust structure
that has stood the tests of the most rigorous experimental scrutinies and the
deepest conceptual criticisms of the past 150 years.
Observation based beginnings, however, have an Achilles' heel.
What if there are escapees from observations that,  if detected, might radically change one's view of Nature.
The question of magnetic monopoles is one such case.
So far, all natural and man-made magnets are found to be dipoles.
And all magnetic field producing electric currents are found to close on themselves and form  loops.  Hence, one has concluded that the magnetic field is divergence free.
But what if one speculates one  magnetic monopole somewhere in the universe,
and what if such speculation is theorized and  expounded on by scientists of Dirac's reputation? Similar questions  could be asked  of the exactness of Coulomb's  inverse square force,  of  the accuracy of  Ampere's,  Faraday's,  and   others' laws.

In what follows we show that there is a reciprocity between
the formal mathematical structure of EM  \& ED on the one hand,
and the universal constancy of the speed of light, on the other.
One implies the other, enabling one to  arrives at an alternative derivation of EM and ED and a different insight.

\section{Minimalist's Electromagnetism}
By the end of the 19th Century the physics community had come to the conclusion
that   light  did not obey the
Galilean law of addition of velocities.
All laboratory and astronomical observations attempting to detect
the motion of light emitting sources, light detecting devices, and a presumed  light propagating medium, Ether, through
experiments using the light itself, yielded negative results.
 Einstein  promoted this conclusion
to the status of an axiom  that
\emph{the speed of light is a universal constant}, the same for all observers.
An immediate corollary to this first principle is the invariance of the
spacetime intervals, that  in inertial frames is expressed as
\begin{equation}
c^2 d\tau^2=c^2dt^2-dx^2-dy^2-dz^2= \eta_{\alpha\beta}dx^\alpha dx^\beta,\label{metric}
\end{equation}
where $\eta_{\alpha\beta}$ is the Minkowski metric tensor.  It will
also be used to lower and raise the vector and tensor indices.

Equation (\ref{metric}) is our departing point.
Assume a test particle of mass $m$ at the spacetime coordinate $x^\gamma$,
4-velocity $U^\alpha(x^\gamma)= dx^\alpha/d\tau$,  and kinetic
4-momentum $p^\alpha(x^\gamma)=m U^\alpha$.
Defined as such, $p^\alpha$ has a constant norm, $|p_{\alpha}p^{\alpha}|^{1/2}=mc$, irrespective of whether
the particle is accelerated or not.
Our next assumption, an everyday observation,  is:
\emph{there are regions of spacetime pervaded by some field in which our assumed  test particles gets accelerated.}
  Hence
\begin{equation}
\f{d p_\alpha}{d\tau}=\frac{\partial p_\alpha}{\partial x^\beta} \frac{d x^\beta}{d\tau}=
\frac{\partial p_\alpha}{\partial x^\beta}U^\beta
=: e F_{\alpha\beta}(x^\gamma) U^\beta \label{4mom}
\end{equation}
where $p_\alpha$ is considered a function of the spacetime coordinates
on particle's orbit  and is differentiated accordingly.
The third equality  is the definition of $F_{\alpha\beta }$,
\begin{eqnarray}
e F_{\alpha\beta } := \partial p_\alpha/\partial x^\beta,\label{deff}
\end{eqnarray}
where $e$, by \emph{ assumption}, is a constant attribute of the test particle and later will be identified as its  electric charge.  Both sides of Eq. (\ref{deff}) are also defined on particle's orbit, specified by some $x^\gamma(\tau)$.
But orbit can be any and every orbit.  Therefore, one is allowed to consider
$F_{\alpha\beta}(x^\gamma)$ a function
of the spacetime coordinates without reference to a specific orbit, and 
identify it with the field responsible for the acceleration of the particle.

The norm of $p^\alpha$ is  constant.
We multiply Eq. (\ref{4mom}) by $p^\alpha$ and find
\begin{equation}
\f{1}{2}\f{d}{d\tau}(p_{\alpha}p^{\alpha)}=\f{e}{m}F_{\alpha\beta}p^{\alpha}p^{\beta}=0.\label{norm}
\end{equation}
Equation (\ref{norm}) implies the antisymmetry of $F_{\alpha\beta}$,
\begin{equation}
F_{\alpha\beta}= -F_{\beta\alpha},~~~ \textrm{tr}  F=0, ~~~F_{\gamma\gamma}=0.\label{fsym}
\end{equation}
In the Appendix we show that a general antisymmetric tensor  can be written as the sum of two other  antisymmetric ones; one of which and the dual of the other are derivable from vector potentials.  Thus
\begin{eqnarray}
F^{\alpha\beta}=  F_1^{\alpha\beta} + {{\cal F}_2}^{\alpha\beta}, ~\textrm{where}~{{\cal F}_2}^{\alpha\beta}= \frac{1}{2} \epsilon^{\alpha\beta\gamma\delta} F_{2\gamma\delta},\label{F12}\\
\textrm{and}~~{F_i}^{\alpha\beta} = \partial ^\beta{ A_i}^\alpha - \partial^\alpha {A_i} ^\beta. ~~i= 1 ~ \& ~ 2.~~~~~~ \label{F12bis}~~~
\end{eqnarray}
Hereafter, the dual of an antisymmetric  tensor denoted by a letter $F$, say, will be shown by the  calligraphic form of the same letter, $\cal F$ here.  Duals are  constructed by  the totally antisymmetric constant pseudotensor $\epsilon^{\alpha\beta\gamma\delta}$ as indicated in Eq. (\ref{F12}).  
   The differential equations for $A_i$ are given in the Appendix,  Eqs. (\ref{waveqn}).
   They are vectors  sourced by the 4-divergences of $F$ and $\cal F$.

 Up to this stage we have discussed kinematics.  Dynamics comes in when one looks for the sources of $F_1$ and $F_2$, and thereof for that of $F$  itself.
 We argue that the field acting on a test particle is generated by the collection of the particles themselves. To find a relation  between the field and the particles \emph{one should look for two similar  characteristics from the field and the particles and equate them}, (our third assumption). 
 
   From the field one may generate two divergence-free 4-vectors:
\begin{eqnarray}
{F^{\alpha\beta}}_{,\beta}={{F_1} ^{\alpha\beta}}_{,\beta}~ \textrm{and}~{{\cal F}^{\alpha\beta}}_{,\beta}= {{F_2}^{\alpha\beta}}_{,\beta}. \label{divfrees}
\end{eqnarray}
In deriving  Eq. (\ref{divfrees}) we have used the fact the 4-divergence of the dual of a tensor derived from a vector potential is zero. 
For the particles, one finds  one and only one  divergence-free 4-vector, namely:
\begin{eqnarray}
J^\alpha(x) \al =\al \sum_n e_n v^\alpha_n({\bf x}_n(t)) \delta^3 ({\bf x} -{\bf  x}_n(t))   \nonumber\\
 \al=\al \sum_n \int e_n{ U_n}^\alpha (x_n(\tau)) \delta^4( x -  x_n(\tau)) d\tau.~~\label{jalpha}
\end{eqnarray}
To construct $J^\alpha$, one assumes a unit 3-volume filled with particles of charges $e_n$, 3-velocities ${\bf v}_n({\bf x}_n(t))$,  4-velocities $U_n^\alpha(\tau)$, and carries out the summation and integration as prescribed above. See e.g. Weinberg  \cite{weinberg} for details.
Evidently, to arrive at the field equations, the only meaningful option is to equate one the 4-vectors of Eq.(\ref{divfrees}) to $J^\alpha$ and equate the other to zero.  We choose the following:
\begin{eqnarray}
{F^{\alpha\beta}}_{,\beta}~\al=\al~ \frac{4\pi}{c}J^\alpha, ~~ ~{J^\alpha}_{,\alpha} = 0,\label{maxwell1}\\
{{\cal F}^{\alpha\beta}}_{,\beta}~\al=\al~ 0.  \label{maxwell2}
\end{eqnarray}

The job is done.   On identifying $F_{\mu\nu}$ with the EM field, and $J^\mu$ with the electric charge-current density of the interacting  particles,  one will recognize Eqs. (\ref{maxwell1}) and (\ref{maxwell2}) as  Maxwell's equations, and Eq. (\ref{4mom})  as  the equations of motion of a particle of the electric charge $e$ under the Lorentz force.   The factor $4\pi/c$ in Eq. (\ref{maxwell1}) is to indicate that we are using the Gaussian units.  There is no point in putting the two  field vectors of Eq. (\ref{divfrees}) proportional to the same $J^\alpha$.    For, one may   always choose an appropriate duality transformation  and bring the transformed equations into the familiar Maxwell's form; see e.g.  \cite{jackson} for this provision.

\subsection{PT Symmetry}
A tacit assumption of at least the classical physics  is  the invariance of   equations of  motion and of fields  under the  space inversion and time reversal (PT symmetry).   To verify the validity of this assumption in the case of  EM field, we first write  Eqs. (\ref{maxwell1}), (\ref{maxwell2}),   and (\ref{4mom}) in their conventional forms in terms of the electric and magnetic vectors.  Let
\begin{eqnarray}
 F^{0i}= - F^{i0} = - E_i,~~
 F^{ij}= \frac{1}{2}\varepsilon^{ijk}B_k. \label{elecmag}
\end{eqnarray}
\begin{eqnarray}
F^{\alpha\beta} = \left[\begin{array}{cccc}
    0&   -E_1&   -E_2&   -E_3\\
  E_1&      0&    B_3&   -B_2\\
  E_2&   -B_3&      0&    B_1\\
  E_3&    B_2&   -B_1&    0
\end{array}\right]. \label{fmunu}
\end{eqnarray}
Equations (\ref{maxwell1}), (\ref{maxwell2}),  and (\ref{4mom}) become
\begin{eqnarray}
\nabla.{\bf E} \al ~=~\al 4\pi\rho, ~~
\nabla\times{\bf B}- \frac{1}{c}\frac{\partial{\bf E}}{\partial t}~=~ \frac{4\pi}{c}{\bf J},\label{ampere}\\
\nabla.{\bf B}\al~=~\al 0, ~~~~~~
\nabla\times{\bf B}- \frac{1}{c}\frac{\partial{\bf E}}{\partial t}~=~0.\label{faraday}\\
 \f{dp^0}{d\tau}\al =\al eF^{0i}U_i, ~~
 \f{dp^i}{d\tau} = e(F^{i0}U_0  + F^{ij}U_j).\label{mom}
\end{eqnarray}
Next we  eliminate $\tau$ in favor of $t$  in Eq. (\ref{mom}),    by letting $d/d\tau= \gamma d/d t$, $U^0=\gamma,~{\bf U}=\gamma {\bf v},~p^0=\gamma m$,  and ${\bf p}=\gamma m {\bf v}$,  where $\gamma=(1-v^2)^{1/2}$.   We obtain.
\begin{equation}
\frac{d}{dt}(\gamma m)=e {\bf E.v},~~\frac{d}{dt}(\gamma m {\bf v})= e(\bf E + v\times B). \label{mombis}
\end{equation}
From Eqs. (\ref{mombis})  one at once concludes the following table of symmetries.
\begin{table}[ht]
%
\begin{tabular}{c|c c c}
                                      &Sp inv              &Time rev     &  Source                                          \\
\hline
{ \bf v}, vector              & odd                & odd            & definition, $ {\bf v}=d{\bf x}/dt$ \\
{\bf E}, vector               & odd                & even          & Eq. (\ref{mombis})                         \\
{\bf B}, pseudovec        &even               & odd            & Eq. (\ref{mombis})                         \\
\end{tabular}\\~~~\\
Symmetries of $\bf E$  are opposite to those of  $\bf B$.
\label{Table:symmetries}
\end{table}
\subsection{Provision for magnetic monopoles}
Since the seminal paper of Dirac \cite{Dirac}, where he  entertains   magnetic monopoles and subsequently concludes  the  quantization of the electric charge, magnetic monopoles have attracted the attention of many great theoretical and experimental physicists.  Of particular importance, beside the  Dirac monopoles that are categorized as QED singularities,  are the parity violating field-theoretic monopoles of 't Hooft - Polyakov  \cite{hooft}.  

From a classical point of view, the fact is that one may speculate a self consisting  EM- and ED- like dynamics in which  a particle may  have both magnetic  and electric  charges,  and  a magnetic  charge-current density may coexist with an electric one, and serve as the source for Eq. (\ref{maxwell2}).   

We argue as follows: The reason for vanishing of the right hand side of    Eqs. (\ref{faraday}), is  the defining Eq. (\ref{deff}), where  we make provision for only a single attribute, $e$, to the test particle and later identify it with its electric charge.  However, from  Eq.(\ref{F12}) and also Eqs. (\ref{F12bis}) and (\ref{vecpot}) below, we now know that  an antisymmetric tensor may in general be written in terms of two  vector potentials.  This makes it possible to go back to  Eq. (\ref{4mom}) and rewrite the test particle  with two attributes $e$ and $g$, say.  Thus,
\begin{equation}
\f{d p^\alpha}{d\tau} = [e {F_1}^{\alpha\beta}+ g {{\cal F}_2}^{\alpha\beta}] U_\beta \label{monopole}.
\end{equation}
One may now construct a magnetic charge-current density, ${J_m}^\alpha$, similar to the electric $J^\alpha$ of Eq. (\ref{jalpha}) with $e_n$ replaced by $g_n$. If different particles or categories of particles have different  $g_n/e_n$ ratios,  then the two vectors   ${J_m}^\alpha$ and  ${J}^\alpha$ will be independent.  This will  allow one to equate them with ${F^{\alpha\beta}}_{,\beta}$ and ${{\cal F}^{\alpha\beta}}_{,\beta}$ of Eq. (\ref{divfrees}),  render the right hand  sides of Eqs. (\ref{maxwell2}) and \ (\ref{faraday}) nonzero, and  make room for magnetic charge-current densities and magnetic  monopoles,  if ever found in Nature.  See Milton et al. \cite{milton} for the resource letter on the theoretical and experimental status of magnetic monopoles.

\section{Summary and Conclusion}

Conventionally,  EM is built on the laboratory findings of Coulomb, Ampere,  Faraday, and  the fact that all magnets found in Nature are dipoles.
To these, Maxwell adds his displacement current to conform with the continuity
of the charge-current density.  To formulate  ED one calls in the Lorentz force law, also an experimentally conceived notion.
These empirical deductions are then promoted to the status of founding principles and EM and ED are formulated.
The universal constancy of $c$ is one of the theoretically derived theorems of the so constructed EM and ED.

Here, we reverse the order of the suppositions and conclusions.
Our founding  principles, also observation based, are:
\begin{description}
\item[] \emph{Speed of light is a universal constant, the first principle of the special theory of relativity.}
\item[]  \emph{There are the so-called charged  particles that mutually interact through a field they themselves create, an everyday observation.}
\end{description}
We find that the spacetime should be pervaded, necessarily,   by a unique rank 2 antisymmetric  tensor, which satisfies Maxwell's equations in  all details,  and the force on a test particle of charge $e$ should necessarily  be the Lorentz force. 

We recall that the pioneering laboratory findings of 18th and 19th Centuries  that led to the formulation of EM and Ed were based on experiments on time independent electrostatic and magnetostatic measurements.  Their  generalization to time dependent circumstances, a bold assumption in its own right, was an  additional  assertion.    Here this assertion  has also emerged as a corollary of the accepted first principles.    

From the first  of  Eqs. (\ref{ampere}),  $\nabla .\textbf{E} = 4\pi \rho$,  one immediately concludes that the Coulomb force
between two charged particles is  exactly $1/r^2$.  (see e.g. \cite{plimpton} for  experimental  verification of Coulomb force).
The same could be said of the exactness of the other empirically accepted laws of EM and ED.

That in the present formalism there is no provision for magnetic monopoles, is because in the equation of motion of the test particle  we  assigned only a single attribute $e$ to the particle.   Had we speculated  particles with two attributes $e$ and $g$ as in Eq. (\ref{monopole}),  we would have made room for magnetic monopoles and magnetic  charge current densities.

 It is  noteworthy that of the two founding principles of the special theory of relativity, namely constant $c$ and same laws of physics in all inertial frames,  only the
fist is used in our formalism.  The invariance of EM and ED in inertial frames
has followed automatically without reference to the second principle. It seems,
at least in the case of EM and ED, the second principle is a conclusion from the first.

Equally noteworthy is the fact that both EM fields and the Lorentz force law emerge as  manifestations of the same set of principles. Together they constitute a whole, whereas in the conventional exposition of EM and ED,  the  Lorentz force law is an  independent assumption from Maxwell's  equations.

  Likewise, the PT  invariance of the EM field is not  independent  from that  of the Lorentz force.  One implies the other,   through  the table of symmetries following  Eq. (\ref{mombis}).

A logician would advise that if $A$ implies $B$ and $B$ implies $A$,  then $A$ and $B$ are equivalent.  Any information contained  in $A$ should also be found in $B$.  Yet it is still thought provoking  how  two simple propositions, constancy of the speed of light and existence of PT observing interacting particles, can lead to a  complex and multi-component structure like EM and ED.

Pedagogics and mnemonics of the formalism is worth noting.   One may derive the whole formalism of  EM and ED   on a hand size piece of paper and memorize it.\\

\section{Appendix}

{ Notation:} Two tensors denoted by the symbol $F$ and its calligraphic form $\cal{ F}$ will be  the dual of each other and  will be connected  as
{ $$  F^{\alpha\beta} = \frac{1}{2} \epsilon^{\alpha\beta\gamma\delta}{\cal{ F}}_{\gamma\delta}, \textrm{vice versa }
 {\cal{F}}^{\alpha\beta} ={ \frac{1}{2}} \epsilon^{\alpha\beta\gamma\delta} F_{\gamma\delta},  $$
  where $ \epsilon^{\alpha\beta\gamma\delta}$ is the totally antisymmetric and  constant  4th rank pseudo-tensor.

{ Remark:}  If an antisymmetric  tensor is derived from a vector potential, its dual will be  divergence free,
$$ {\cal  F}^{\alpha\beta}_{~~~,\beta} = \frac{1}{2} \epsilon^{\alpha\beta\gamma\delta}(\partial_\delta A_\gamma - \partial_\gamma A_\delta)_{,\beta}= 0.$$,

{\bf Theorem.}  Any antisymmetric tensor $F$ can be written as the sum of two other antisymmetric tensors, $F_1$ and dual$F_2$, where both $F_1$  and $F_2$ are derived from vector potentials, sourced  by divergences of $F$ and $\cal F$, respectively.   Thus,
\begin{equation}
F^{\alpha\beta} = F_1^{\alpha\beta} + {\cal F}_2^{\alpha\beta},  ~\textrm{and }~
{\cal F}^{\alpha\beta} ={\cal  F}_1^{\alpha\beta} +  F_2^{\alpha\beta},  \label{F12bis}
\end{equation}
where
   \begin{equation}
  F_1^{\alpha\beta} = \partial ^\beta A_1^\alpha - \partial^\alpha A_1 ^\beta, \textrm{~and~}
F_2^{\alpha\beta} = \partial ^\beta A_2^\alpha - \partial^\alpha A_2 ^\beta. \label{vecpot}
\end{equation}
One has the gauge freedom to choose $A$'s divergence free.  Now substituting Eqs. (\ref{vecpot}) in Eqs. (\ref{F12bis}) and taking their 4-divergence  gives
\begin{eqnarray}
\partial_\beta \partial^\beta A_1^{~\alpha} \al = \al F^{\alpha\beta}_{~~~,\beta} ={ { F_1}^{\alpha\beta}}_{,\beta}, ~~~\partial_\alpha {A_1}^\alpha = 0, \nonumber
\\
\partial_\beta \partial^\beta A_2^{~\alpha} \al = \al{\cal F}^{\alpha\beta}_{~~~,\beta} ={ { F_2}^{\alpha\beta}}_{,\beta},~~~\partial_\alpha {A_2} ^\alpha = 0. \label{waveqn}
\end{eqnarray}
Equations (\ref{waveqn}) are two wave equations sourced by the 4-divergences  of  $F$ and its dual, $\cal F$.
Their retarded causal  solutions are the soughtafter vector potentials. 
To prove the theorem it is sufficient to substitute these   retarded solutions in Eq. (\ref{vecpot}),  then the results in Eq. (\ref{F12bis}) and obtain  the same $F$ that one had started with.  Calculations are extensive but straightforward.

{\bf Acknowledgment:}
The Author wishes to thank M. R. Khajehpour,  B. Mashhoon, V. Karimipour, S. Shiekh Jabbari, A. R. Valizade, and B. Farnudi for fruitful discussions and suggestions.


\begin{thebibliography}{00}

  \bibitem{weinberg}
 Weinberg, S., Gravitation and Cosmology: Principles and applications of the general theory of gravity,  pp. 40 - 46, John Wiley \& Sons, New York, 1972
 
  \bibitem{jackson}
Jackson, J. D., Classical Electrodynamics, John Wiely, New York, third edition, p. 274, 1999.
 
 \bibitem{Dirac}
  Dirac,  P. A. M.,  Proc. R. Soc. London, \textbf{A 133}, 1931;  Phys. Rev. \textbf{74},   1948.

    \bibitem{hooft}
 't Hooft, G., Nucl. Phys. {\bf  B 79}, 276, 1974.\\
 %
 Polyakov, A. M.. JETP Lett. {\bf 20} 194, 1974,  and  JETP {\bf  41}, 988, 1975.

 \bibitem{milton}
 Milton, K. A., arXiv:hep-ex10602040.
 
  \bibitem{plimpton}
  Plimpton, S. J.; Lawton, W. E., Phys. Rev. \textbf{50}, 1066, 1936.

\end{thebibliography}
\end{document}